\documentclass[11pt]{article}
\usepackage{amssymb}
\usepackage{amsmath}
\usepackage{amsfonts}
\usepackage{amsbsy}
\usepackage[dvips]{graphicx}
\usepackage{color}

\newcommand{\bfga}{\boldsymbol{\gamma}}

\newcommand{\bfp}{\boldsymbol{p}}
\newcommand{\bfx}{\boldsymbol{x}}
\begin{document}
\font\titlefont=cmbx10 scaled\magstep1
\title{BCS theory of hadronic matter at high densities}
\author{
Henrik Bohr\\
{\it \small Department of Physics, B.307, Danish Technical
University,}\\{ \it \small DK-2800 Lyngby, Denmark}\\Prafulla K.
Panda\\{\it\small Departmenet of Physics, C.V. Raman College of
Engineering,}\\{ \it \small  Vidya Nagar, Bhubaneswar-752054, India}
\\Constan\c{c}a Provid\^encia,
Jo\~ao da Provid\^encia\\
{\it \small CFC, Departamento de F\'\i sica, Universidade de Coimbra,}\\
{\it \small  P-3004-516 Coimbra, Portugal} } \maketitle
\abstract{ The equilibrium between 
the so-called 2SC and CFL phases of strange quark matter at high
densities is investigated in the framework of a simple schematic
model of the NJL type. Equal densities are assumed for quarks $u,d$
and $s$. The 2SC phase is here described by a color-flavor symmetric
state, in which the quark numbers are independent of the
color-flavor combination. In the CFL phase the quark numbers depend
on the color-flavor combination, that is, the number of quarks
associated with the color-flavor combinations $ur,dg,sb$ is
different from the number of quarks associated with the color flavor
combinations $ug,ub,dr,db,sr,sg$. We find that the 2SC phase is
stable for a chemical potential $\mu$ below $\mu_c=0.505$ GeV, while
the CFL phase is stable above, the equilibrium pressure being
$P_c=0.003$ GeV$^4$. We have used a 3-momentum regularizing cutoff
$\Lambda=0.8$ GeV, which is somewhat larger than is  usual in NJL
type models. This should be adequate if the relevant chemical
potential does not exceed 0.6 GeV. }
\def\ve{\varepsilon}
\def\e{{\rm e}}
\def\bfp{{\bf p}}
\def\d{{\rm d}}
\def\ol{\overline}
\def\bfx{{\bf x}}
\def\<{\langle}
\def\>{\rangle}
\def\ve{\varepsilon}
\bigskip
\bigskip
\section{Introduction}
The color superconducting phase in quark matter is a topic of great
current interest which has attracted the attention of many authors
\cite{alford}. It is expected to be realized at very high densities,
several times the equilibrium density of nuclear matter. For a
recent review, see \cite{alford1}. In the literature, the color
superconducting phase is described by a BCS state which breaks the
SU(3) color symmetry. This is unavoidable. However, in the
conventional treatment of color superconductivity, a BCS state
vector is used which gives rise to color-dependent quark numbers.
This is undesirable and can be easily avoided \cite{bohr1}. Rudolf
Peierls \cite{peierls} classifies the BCS approach as a {\it
symmetry breaking approximation} meaning that the relevant symmetry
is not globally broken by the physical state itself but by the
approximation used to take into account important correlations which
are present in the physical state. In this note we make use of a
version of the BCS formalism for quark matter \cite{bohr1} which, as
far as possible, is compatible with color symmetry. Here, strange
quark matter and some of its phases is investigated. In the
conventional 2SC phase, 2 specified colors and 2 specified flavors
are paired, the third color and flavor remaining as spectators. In
addition, in the 2SC phase, beyond the chemical potential $\mu$
which fixes the total number of quarks, an extra Lagrange multiplier
$\mu_8$ is used to insure that the number of quarks is color
independent, or, equivalently, to insure that
$\langle\lambda_8\rangle=0$  \cite{buballa,ruster}. If an extra
constraint is used,  the thermodynamical potential is increased, the
pressure is lowered and so the stability reduced, which is
undesirable. We propose an alternative approach \cite{bohr1} in
which the condition $\langle\lambda_8\rangle=0$ is automatically
achieved without imposing it as an additional constraint as is done,
for instance in \cite{ruster}. Our approach has the advantage of
being elementary and leads to several possible phases, namely, a
color-flavor symmetric phase, a color symmetric flavor asymmetric
phase, a color-flavor asymmetric phase. These phases are essentially
equivalent to the 2SC phase except that they do not involve the
Lagrange multiplier $\mu_8$ and are therefore more stable. The
so-called color flavor locked (CFL) phase \cite{alford2} is also
described. It is presently accepted that quark and gluon fields are
the building blocks of hadronic matter, in the framework of quantum
chromodynamics (QCD). A diversity of phases is expected at high
densities: chiral-symmetry restoration, deconfinement and
color-superconductivity. We focus on high densities for which chiral
symmetry is expected to have been restored, that is, phases on which
the color condensate $\langle \bar q q \rangle$ is expected to be no
longer present. Since, due to the complexity of the theory, it is
extremely difficult, if not impossible, to obtain exact results
directly from QCD when perturbation theory cannot be applied, we
follow the common practice and resort to an effective model of the
Nambu-Jona-Lasinio (NJL) type containing a QCD inspired four-fermion
interaction \cite{NJL,klevansky,hatsuda,buballa,koide}. The present
calculation is admittedly very schematic. In the end, we assume that
the current masses of all quarks, including strange quarks, may be
neglected. This  rough approximation is acceptable if the chemical
potential $\mu$ is high enough. The phase diagram of high density
hadronic matter has been realistically treated in \cite{ruster}.
Although our discussion is based on a schematic and qualitative
model, it suggests, for the critical chemical potential, a
transition from the color symmetric flavor asymmetric phase to the
CFL phase. In order to account for the tails of the occupation
numbers, we use a 3-momentum regularizing cutoff $\Lambda=0.8$ GeV,
which is somewhat larger than is  usual in NJL type models. We find
that the color-flavor symmetrical phase is stable below the critical
density $\rho_1=0.022$ GeV$^3$. The CFL phase is stable above the
critical density $\rho_2=0.025$ GeV$^3$. Between $\rho_1$ and
$\rho_2$ both phases co-exist.

The properties of quark matter at high densities may be of interest
in connection with the description of the interior of neutron stars.
In \cite{alford2} it is suggested that it is made of strange matter,
at least if the density is sufficiently high. The present results
are consistent with this hypothesis.

\begin{figure}[ht]
\vspace{1.5cm} \centering
\includegraphics[width=0.75\linewidth,angle=0]{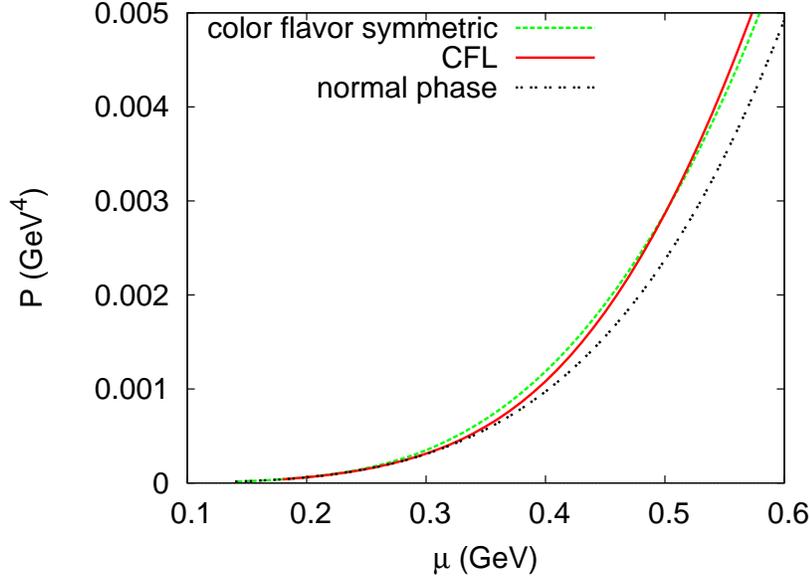}
\caption{Transition from the color-flavor symmetric phase to the CFL
phase. This figure shows that the color flavor symmetric phase is
stable for $\mu$ below $\mu_c=0.505$ GeV, and the CFL phase is
stable for $\mu$ above that chemical potential, the equilibrium
pressure being $P_c=0.003$ GeV$^4$. The curves for the color
symmetric flavor asymmetric and for the color-flavor asymmetric
phases are superimposed with the curve for the color-flavor
symmetric phase, which is shown.
} \label{exci4}
\end{figure}
\section{Hamiltonian and thermodynamical potential} The
thermodynamical potential is the Hamiltonian  constrained by the
Fermion number conservation. The chemical potential $\mu$ is the
Lagrange multiplier which fixes the number of quarks. In the
mean-field approximation, the thermodynamical potential operator
reads,
\begin{eqnarray*}\hat
K_{MFA}=\int\d^3{\bfx}\left[\overline\psi(\bfp\cdot\bfga+M-\mu\gamma_0)\psi+{1\over2}\sum_{j,a\in\{2,5,7\}}
\left((\Delta^*_{ja}\overline\psi^Ci\gamma_5\lambda^{c}_j\lambda^{f}_a\psi+h.c.)+{|\Delta_{ja}|^2\over2G_c}\right)\right],
\end{eqnarray*}
where
$$\Delta_{ja}=-G_c\langle\overline\psi^Ci\gamma_5\lambda^{c}_j\lambda^{f}_a\psi\rangle.$$ Here,
$\langle\overline\psi^Ci\gamma_5\lambda^{c}_j\lambda^{f}_a\psi\rangle$
denotes the expectation value of  the diquark condensate $(
\overline\psi^Ci\gamma_5\lambda^{c}_j\lambda^{f}_a\psi)$ in the BCS
vacuum, and $\lambda_j^{c},\lambda_a^{f},\,j,a\in\{2,5,7\}$ are the
antisymmetric Gell-Mann matrices, for color and flavor,
respectively. The current quark mass is $M$ and the chemical
potential is $\mu$.  We wish to work in momentum space, so we need
the expression in momentum space of the operators
$(\bar\psi^Ci\gamma_5\lambda_j^{c}\lambda_a^{f}\psi)$ for
$j,a\in\{2,5,7\}$. These quantities are associated with the
$\overline 3$ representations of color and flavor $SU(3)$. The mean
field Hamiltonian density is expressed in terms of the diquark
condensate.
However, the
mean field Hamiltonian itself involves the respective integral
$
\int\d^3\bfx(\bar\psi^Ci\gamma_5\lambda_j^{c}\lambda_a^{f}\psi).$
For $a=2,j=2$, we find \begin{eqnarray*}&&{1\over
V}\int\d^3\bfx(\bar\psi^Ci\gamma_5\lambda_2^{c}\lambda_2^{f}\psi)=
\sum_p(c_{2dp}c_{1u\overline p}+c_{1up}c_{2d\overline
p}-c_{2up}c_{1d\overline p}-c_{1dp}c_{2u\overline p} )\eta_p\\&&
\quad\quad\quad\quad\quad\quad\quad\quad+\sum_p(\tilde
c_{2dp}^\dagger \tilde c_{1u\overline p}^\dagger +\tilde
c_{1up}^\dagger \tilde c_{2d\overline p}^\dagger -\tilde
c_{2up}^\dagger \tilde c_{1d\overline p}^\dagger -\tilde
c_{1dp}^\dagger \tilde c_{2u\overline p}^\dagger )\eta_p,\quad
\eta_p=-\eta_{\overline p},\end{eqnarray*} with analogous
expressions for the other choices of $a,j$. Here, $c_{jbp}$
annihilates a quark with color $j$ flavor $b$ and momentum-helicity
$p$, while $\tilde c_{jbp}^\dagger$ creates an anti-quark with color
$j$ flavor $b$ and momentum-helicity $p$, and $\overline p$ is the
state with the same helicity as $p$ but opposite momentum. Since the
Cooper instability occurs at the Fermi surface, we will ignore
anti-quarks, in the sequel. Therefore, we will assume that in the
mean-field approximation the thermodynamical potential reads simply,
\begin{eqnarray}\hat
K_{MFA}&=&\sum_{iap}\epsilon_pc^\dagger_{iap}c_{iap}\nonumber\\
&+&\sum_{ijkabcp}( \Delta^*_{kc} ~{c_{jbp}~c_{ia\overline
p}}~\epsilon_{ijk}\epsilon_{abc}\eta_p+ \Delta_{kc}
~c^\dagger_{iap}~c^\dagger_{jb\overline
p}~\epsilon_{ijk}\epsilon_{abc}\eta_p)+{V\over
2G_c}\sum_{jc}\Delta_{jc}\Delta_{jc}^*,\nonumber\\\Delta_{kc}&=&{G_c\over
V}\sum_{ijabp}\langle {c_{jbp}~c_{ia\overline
p}}\rangle~\epsilon_{ijk}\epsilon_{abc}\eta_p,\label{KMFA}
\end{eqnarray}where $\epsilon_p=\sqrt{p^2+M^2}-\mu$.
It is convenient to define the effective gap as
$$\Delta_{eff}=\sqrt{{1\over9}~\sum_{jc}\Delta_{jc}\Delta_{jc}^*}~.$$
 We denote the color
indices as $1,2,3$,  meaning {\it red, green, blue}, and the flavor
indices as $a,b,c$. Upon a color or flavor rotation the gaps
$\Delta_{jc}$, $j\in\{1,2,3\},~c\in\{u,d,s\}$, change but
$\Delta_{eff}$ and the equation of state do not change.
\begin{figure}[ht]
\vspace{1.5cm} \centering
\includegraphics[width=0.75\linewidth,angle=0]{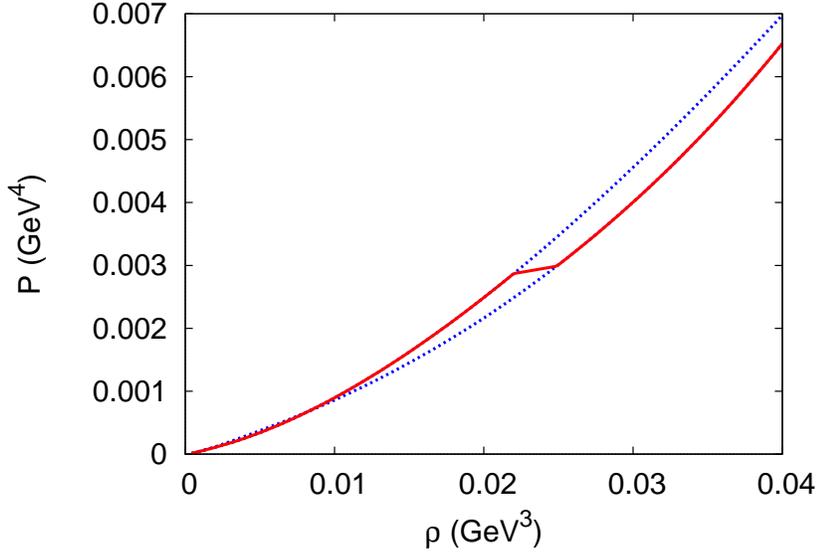} 
\caption{Pressure vs  quark density. This figure illustrates the
phase equilibrium between the CFL phase (at higher densities) and
color-flavor symmetric phase (at lower densities). The dotted curve
shows the un-physical continuation of the CFL curve (at lower
densities) and of the color flavor symmetric curve (at higher
densities). The two phases coexist for $\rho$ between $\rho_1=0.022$
GeV$^3$ and $\rho_2=0.025$ GeV$^3$, the equilibrium pressure being
$P_c=0.003$ GeV$^4$. The curves for the color symmetric flavor
asymmetric and for the color-flavor asymmetric phases are
superimposed with the curve for the color-flavor symmetric phase,
which is shown.  The maximum density presented corresponds to 9.808
times the nuclear saturation density.} \label{exci1}
\end{figure}
\section{Color superconductivity}
We consider the following super-conducting phases: 1) a color-flavor
symmetric phase; 2) a color symmetric, flavor asymmetric phase; 3) a
color-flavor asymmetric phase; 4) the so-called CFL phase
\cite{alford2}. The color-flavor symmetric phase is defined by the
condition that the gaps $\Delta_{ia}$  are non-zero and equal,
independently of color and flavor, that is,
$\Delta_{ia}=\Delta\neq0,$ $i\in\{1,2,3\}$, $a\in\{u,d,s\}$. The
color symmetric flavor asymmetric phase is defined by
$\Delta_{iu}=\Delta_{id}=\Delta,~\Delta_{is}=0,~i\in\{1,2,3\}$. The
color and flavor asymmetric phase  is defined by
$\Delta_{1u}=\Delta_{2u}=\Delta_{1d}=\Delta_{2d}=\Delta,~\Delta_{is}=\Delta_{3a}=0,~i\in\{1,2,3\},~a\in\{u,d,s\}$.
In the CFL phase, by definition,  the gaps satisfy
$\Delta_{1u}=\Delta_{2d}=\Delta_{3s}\neq 0,$ while
$\Delta_{1d}=\Delta_{1s}=\Delta_{2u}=\Delta_{2s}=\Delta_{3u}=\Delta_{3d}=0.$
\subsection{Color-flavor symmetric phase}
The BCS vacuum of the color-flavor symmetric phase is given by
$$|\Phi\rangle=\e^S|\Phi_0\rangle,\quad|\Phi_0\rangle=\prod_{iap(\epsilon_p\leq0)}c^\dagger_{iap}|0\rangle,$$
where
\begin{eqnarray*}S=-\sum_{p(\epsilon_p>0)}K_p\sum_{ijkabc}{c_{iap}^\dagger}{c_{jb\overline
p}^\dagger}\epsilon_{ijk}\epsilon_{abc}\eta_p
-\sum_{p(\epsilon_p\leq0)}\widetilde
K_p\sum_{ijkabc}c_{iap}c_{jb\overline
p}\epsilon_{ijk}\epsilon_{abc}\eta_p.
\end{eqnarray*}
It follows that the BCS vacuum satisfies  
$$d_{iap}|\Phi\rangle=d_{ia\overline p}|\Phi\rangle=0,\quad
i\in\{1,2,3\},a\in\{u,d,s\}$$ being, for
$\epsilon_p=\sqrt{p^2+M^2}-\mu> 0$,
$$d_{1up}=c_{1up}+K_p(c^\dagger_{2d\overline{p}}-c^\dagger_{2s\overline{p}}-
c^\dagger_{3d\overline{p}}+c^\dagger_{3s\overline{p}})\eta_p,\quad\eta_{\overline
p}=-\eta_p,
$$
the other operators are obtained by cyclic permutations of the
indices 1,2,3 and $u,d,s$, while  for
$\epsilon_p=\sqrt{p^2+M^2}-\mu\leq 0$
$$d_{1up}=c_{1up}^\dagger-\widetilde K_p(c_{2d\overline{p}}-c_{2s\overline{p}}-
c_{3d\overline{p}}+c_{3s\overline{p}})\eta_p,\quad 
$$ plus cyclic permutations of the indices 1,2,3 and $u,d,s$. The
operators $d_{iap},~d_{iap}^\dagger$do not obey canonical
anti-commutation relations, since,
$\{d_{iap},~d_{jbq}^\dagger\}\neq\delta_{ij}\delta_{ab}\delta_{pq}$,
although $\{d_{iap},~d_{jbq}\}=0 .$ New operators $f_{iap}$ are
easily constructed as linear combinations of the operators $d_{iap}$
which obey the  canonical anti-commutation relations. These
operators are not needed for the present purpose, so that their
expressions are not presented.
It is easily found that the optimal BCS vacuum leads to the gap
equation
$$1={2G\over V}\left(\sum_{p,\epsilon_p\leq0}+\sum_{p,\epsilon_p>0}\right){1\over\sqrt{\epsilon_p^2+9\Delta^2}},$$
and to the expression for the average thermodynamical potential
(\ref{KMFA}), which reads
\begin{eqnarray*}&&\langle \hat
K_{MFA}\rangle=\sum_{p,\epsilon_p\leq0}\left(7\epsilon_p-2\sqrt{\epsilon_p^2+9\Delta^2}
\right)+\sum_{p,\epsilon_p>0}\left(2\epsilon_p-2\sqrt{\epsilon_p^2+9\Delta^2}
\right)+9V{\Delta^2\over2G_c}.
\end{eqnarray*}
The equations of motion of the operators $c^\dagger_{iap}$ are
straightforwardly obtained. As an example, the following commutation
relation is presented,
$$[\hat K_{MFA},c^\dagger_{1up}]=\epsilon_pc^\dagger_{1up}-\Delta(c_{2d\overline p}+c_{3s\overline p}-c_{2s\overline p}-c_{3d\overline
p}).$$ Using these commutation relations, standard techniques allow
the determination of linear combinations $f_p^\dagger=\sum
(x_{ja}c^\dagger_{jap}+y_{ja}c_{ja\overline p})$ satisfying
$$[\hat K_{MFA},f_p^\dagger]=\omega_pf_p^\dagger.$$
The diagonalization of the operator $\hat K_{MFA}$ is now easily
achieved. One finds that the eigenvalues $\omega_p$ are $\epsilon_p$
(5-fold degenerate) and $\sqrt{\epsilon_p^2+9\Delta^2}$ (4-fold
degenerate). Finally, for appropriate fermion operators
$f_{kp}^\dagger$ we may write
$$\hat K_{MFA}=\sum_{j=1}^5|\epsilon_p| f_{jp}^\dagger f_{jp}+\sum_{j=6}^9\sqrt{\epsilon_p^2+9\Delta^2} f_{jp}^\dagger f_{jp}+
\langle \hat K_{MFA}\rangle.$$ In the color-flavor symmetric case,
we have $\Delta_{eff}=\Delta.$  We observe that in the color-flavor
symmetric phase the BCS quasi particles $f_{kp}^\dagger$ are not
characterized by well defined quantum numbers. This is the price we
have to pay for obtaining color-flavor independent quark numbers.
\begin{figure}[ht]
\vspace{1.5cm} \centering
\includegraphics[width=0.75\linewidth,angle=0]{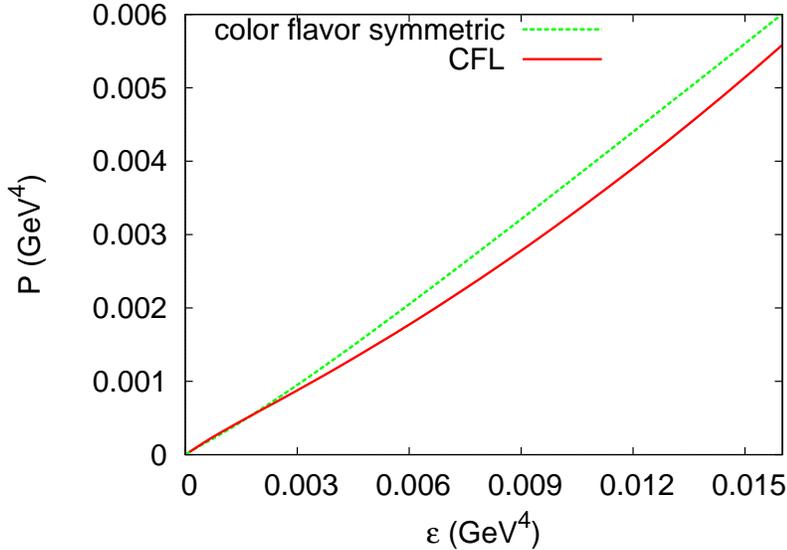} 
\caption{ Pressure  vs energy density. The red curve is physically
meaningful for $P>0.003$ GeV. The other curve is physically
meaningful below that pressure. The curves  for the color symmetric
flavor asymmetric and for the color-flavor asymmetric phases are
superimposed with the curve for the color-flavor symmetric phase,
which is shown.} \label{exci6}
\end{figure}

\begin{figure}[ht]
\vspace{1.5cm} \centering
\includegraphics[width=0.75\linewidth,angle=0]{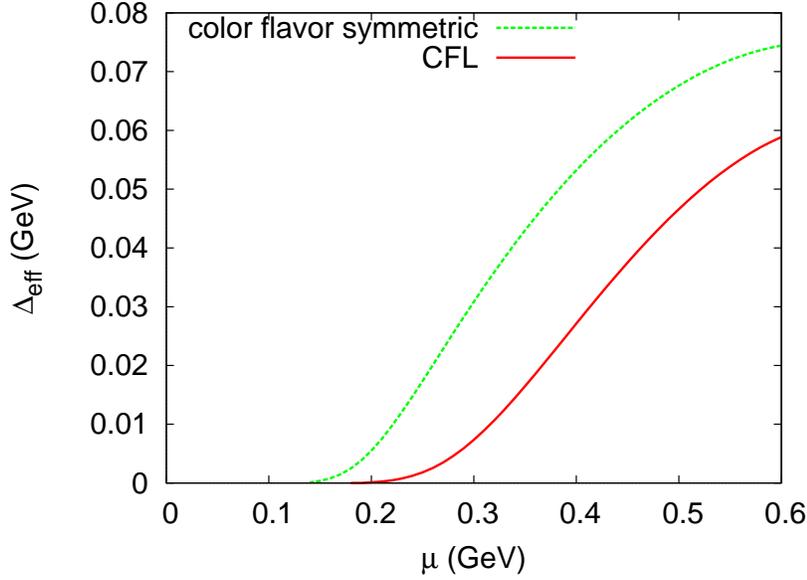} 
\caption{The efective gap $\Delta_{eff}$ for several phases.
 The curves  for the color symmetric
flavor asymmetric and for the color-flavor asymmetric phases are
superimposed with the curve for the color-flavor symmetric phase,
which is shown.} \label{exci7}
\end{figure}
\subsection{CFL phase} Next, we discuss the so-called CFL
phase \cite{alford2}.

 For a fixed $p$, let us consider the following effective
Hamiltonian,
\begin{eqnarray*}H_{eff}&=&\epsilon\sum_{i=1}^3\sum_{a\in\{u,d,s\}}(c^\dagger_{iap}c_{iap}+c^\dagger_{ia\overline p}c_{ia\overline
p})\\&
+&\Delta\sum_{i,j,k=1}^3\sum_{a,b\in\{u,d,s\}}(~{c_{jbp}~c_{ia\overline
p}}~
+ ~c^\dagger_{iap}~c^\dagger_{jb\overline
p})~\epsilon_{ijk}\epsilon_{abc_k},
\end{eqnarray*} where $c_1=u,~c_2=d,~c_3=s,$
which involves only the di-quark condensates $
(c_{2dp}c_{1u\overline p}+c_{1up}c_{2d\overline
p}-c_{2up}c_{1d\overline p}-c_{1dp}c_{2u\overline p} ),
 $ $ (c_{3sp}c_{2d\overline p}+c_{2dp}c_{3s\overline
p}-c_{3dp}c_{2s\overline p}-c_{2sp}c_{3d\overline p} ),
 $ $ (c_{1up}c_{3s\overline p}+c_{3sp}c_{1u\overline
p}-c_{1sp}c_{3u\overline p}-c_{3up}c_{1s\overline p} ),$ and their
hermitian adjoints.
Having in mind the equations of motion of the operators
$c^\dagger_{iap}$, exemplified by
\begin{eqnarray*}&&[H_{eff},c^\dagger_{1up}]=\epsilon c^\dagger_{1up}-\Delta(c_{2d\overline p}+c_{3s\overline
p}),\\&&[H_{eff},c^\dagger_{1dp}]=\epsilon
c^\dagger_{1dp}+\Delta(c_{2u\overline p}),\quad
[H_{eff},c^\dagger_{1sp}]=\epsilon
c^\dagger_{1sp}+\Delta(c_{3u\overline p}),
\end{eqnarray*} we easily show that $H_{eff}$ is diagonalized by the BCS
transformation
\begin{eqnarray}&&d_{jp}^\dagger=\sum_{i=1}^3(x^{(j)}_{i}c^\dagger_{ia_ip}+y^{(j)}_{i}c_{ia_i\overline
p}),\quad d_{j\overline
p}^\dagger=\sum_{i=1}^3(x^{(j)}_{i}c^\dagger_{ia_i\overline
p}-y^{(j)}_{i}c_{ia_i p}), \quad
j=1,2,3,\quad\label{bcsCFL}\end{eqnarray} being
$a_1=u,~a_2=d,~a_3=s$, \color{black} and
\begin{eqnarray*}&&x_1^{(1)}=x_2^{(1)}=x_3^{(1)}=\sqrt{{\epsilon+\sqrt{\epsilon^2+4\Delta^2}\over{6
\sqrt{\epsilon^2+4\Delta^2}}}},\quad
y_1^{(1)}=y_2^{(1)}=y_3^{(1)}=\sqrt{{-\epsilon+\sqrt{\epsilon^2+4\Delta^2}\over{6
\sqrt{\epsilon^2+4\Delta^2}}}},\\&& 
x_1^{(2)}=-x_2^{(2)}
={1\over2}\sqrt{{\epsilon+\sqrt{\epsilon^2+\Delta^2}\over{
\sqrt{\epsilon^2+\Delta^2}}}},
\quad x_1^{(3)}=x_2^{(3)}
=-{1\over2}x_3^{(3)}
={1\over2}\sqrt{{\epsilon+\sqrt{\epsilon^2+\Delta^2}\over{3
\sqrt{\epsilon^2+\Delta^2}}}},\\&&
-y_1^{(2)}=y_2^{(2)}={1\over2}\sqrt{{-\epsilon+\sqrt{\epsilon^2+\Delta^2}\over{
\sqrt{\epsilon^2+\Delta^2}}}},
\quad
-y_1^{(3)}=-y_2^{(3)}={1\over2}y_3^{(3)}
={1\over2}\sqrt{{-\epsilon+\sqrt{\epsilon^2+\Delta^2}\over{3
\sqrt{\epsilon^2+\Delta^2}}}},\\&&
x_3^{(2)}=y_3^{(2)}
=0.\end{eqnarray*}The relations (\ref{bcsCFL}) are completed
by\begin{eqnarray*}&&d_{4p}^\dagger=Xc_{1dp}^\dagger+Yc_{2u\overline{p}},\quad
d_{5p}^\dagger=Xc_{2up}^\dagger+Yc_{1d\overline{p}},\\&&
d_{4\overline p}\dagger=Xc_{1d\overline p}^\dagger-Yc_{2u{p}},\quad
d_{5\overline p}\dagger=Xc_{2u\overline p}^\dagger-Yc_{1d{p}},
\end{eqnarray*}plus analogous relations for the remaining color-flavor
combinations, being
\begin{eqnarray*}
X=\sqrt{{\epsilon+\sqrt{\epsilon^2+\Delta^2}\over{2
\sqrt{\epsilon^2+\Delta^2}}}},
\quad Y =\sqrt{{-\epsilon+\sqrt{\epsilon^2+\Delta^2}\over{2
\sqrt{\epsilon^2+\Delta^2}}}}.\end{eqnarray*} Its eigenvalues are,
$\sqrt{\epsilon^2+4\Delta^2}~,$ (2-fold degenerate) and
$\sqrt{\epsilon^2+\Delta^2}~$ ({16}-fold degenerate), so that
\begin{eqnarray*}H_{eff}&=&\sqrt{\epsilon^2+4\Delta^2}(d^\dagger_{1p}d_{1p}+d^\dagger_{1\overline
p}d_{1\overline p}) +\sqrt{\epsilon^2+\Delta^2}
~{\sum_{j=2}^9(d^\dagger_{jp}d_{jp}+d^\dagger_{j\overline p}d_{j
\overline
p})}\\&+&{9\epsilon-8\sqrt{\epsilon^2+\Delta^2}-\sqrt{\epsilon^2+4\Delta^2}}.\end{eqnarray*}
We postulate that the CFL phase is the vacuum of the quasi-particles
defined by (\ref{bcsCFL}) for $p$ such that $\epsilon_p>0$ and by a
similar transformation, in which, however, the states
$c^\dagger_{jap}$ are regarded as antipaticles, for $p$ such that
$\epsilon_p\leq 0$, that is

\begin{eqnarray}&&d_{jp}=\sum_{i=1}^3(x^{(j)}_{i}c^\dagger_{ia_ip}+y^{(j)}_{i}c_{ia_i\overline
p}),\quad d_{j\overline
p}^\dagger=\sum_{i=1}^3(x^{(j)}_{i}c^\dagger_{ia_i\overline
p}-y^{(j)}_{i}c_{ia_i p}), \label{bcsCFLholes}\end{eqnarray}
The relations (\ref{bcsCFLholes}) are completed
by\begin{eqnarray*}&&d_{4p}=Xc_{1dp}^\dagger+Yc_{2u\overline{p}},\quad
d_{5p}=Xc_{2up}^\dagger+Yc_{1d\overline{p}},\\&& d_{4\overline
p}=Xc_{1d\overline p}^\dagger-Yc_{2u{p}},\quad d_{5\overline
p}=Xc_{2u\overline p}^\dagger-Yc_{1d{p}},
\end{eqnarray*}plus analogous relations for the remaining color-flavor
combinations.
 For the gap equation in the CFL phase we obtain
$$1={G\over 3V}\left(\sum_{p,\epsilon_p\leq0}+\sum_{p,\epsilon_p>0}\right)\left({4\over\sqrt{\epsilon_p^2+4\Delta^2}}
+{8\over\sqrt{\epsilon_p^2+\Delta^2}}\right)$$  and for the average
thermodynamical potential we find
\begin{eqnarray*}&&\langle \hat
K_{MFA}\rangle=\sum_{p,\epsilon_p\leq0}\left(-\sqrt{\epsilon_p^2+4\Delta^2}-8\sqrt{\epsilon_p^2+\Delta^2}
\right)\\&&+\sum_{p,\epsilon_p>0}\left(9\epsilon_p-\sqrt{\epsilon_p^2+4\Delta^2}-8\sqrt{\epsilon_p^2+\Delta^2}
\right)+3V{\Delta^2\over2G_c}.
\end{eqnarray*}  We observe that the gap equation for the CFL phase
is essentially different from the gap equations for all  the other
phases. In the CFL phase, $\Delta_{eff}=\Delta/\sqrt{3}.$
\subsection{Color symmetric flavor asymmetric phase}
In the case of the color symmetric flavor asymmetric phase, for
$\epsilon_p=\sqrt{p^2+M^2}-\mu\geq 0$, the BCS vacuum is annihilated
by the operators
$$d_{1up}=c_{1up}+K_p(c^\dagger_{2d\overline p}-c^\dagger_{3d\overline p}),
\quad d_{1dp}=c_{1dp}+K_p(c^\dagger_{2u\overline
p}-c^\dagger_{3u\overline p}),
$$
$$
d_{2up}=c_{2up}+K_p(c^\dagger_{3d\overline p}-c^\dagger_{1d\overline
p}),\quad d_{2dp}=c_{2dp}+K_p(c^\dagger_{3u\overline
p}-c^\dagger_{1u\overline p})
$$
$$d_{3up} =c_{3up}+K_p(c^\dagger_{1d\overline p
}-c^\dagger_{2d\overline p}),\quad d_{3dp}
=c_{3dp}+K_p(c^\dagger_{1u\overline p }-c^\dagger_{2u\overline p})$$
$$d_{isp}=c_{isp},\quad i\in\{1,2,3\}$$
while for $\epsilon_p=\sqrt{p^2+M^2}-\mu\leq 0$, the BCS vacuum is
annihilated by
$$d_{1up}=c_{1up}^\dagger-\widetilde K_p(c_{2d\overline p}-c_{3d\overline p}),
\quad d_{1dp}=c_{1dp}^\dagger-\widetilde K_p(c_{2u\overline
p}-c_{3u\overline p}),
$$
$$
d_{2up}=c_{2up}^\dagger-\widetilde K_p(c_{3d\overline
p}-c_{1d\overline p}),\quad d_{2dp}=c_{2dp}^\dagger-\widetilde
K_p(c_{3u\overline p}-c_{1u\overline p})
$$
$$
d_{3up}^\dagger=c_{3up}^\dagger-\widetilde K_p(c_{1d\overline p
}-c_{2d\overline p }),\quad
d_{3dp}^\dagger=c_{3dp}^\dagger-\widetilde K_p(c_{1u\overline p
}-c_{2u\overline p })$$
$$d_{isp}=c_{isp}^\dagger,\quad i\in\{1,2,3\}.$$
In this case, the gap equation reads,
$$1={2G\over V}\left(\sum_{p,\epsilon_p\leq0}+\sum_{p,\epsilon_p>0}\right){1\over\sqrt{\epsilon_p^2+3\Delta^2}}.$$
and the 
thermodynamical potential reads,
\begin{eqnarray*}&&\langle \hat
K_{MFA}\rangle=\sum_{p,\epsilon_p\leq0}\left(7\epsilon_p-2\sqrt{\epsilon_p^2+3\Delta^2}
\right)+\sum_{p,\epsilon_p>0}\left(2\epsilon_p-2\sqrt{\epsilon_p^2+3\Delta^2}
\right)+3V{\Delta^2\over2G_c}.
\end{eqnarray*}
In the color symmetric flavor asymmetric phase,
$\Delta_{eff}=\Delta/\sqrt{3}.$
\begin{figure}[ht]
\vspace{1.5cm} \centering
\includegraphics[width=0.75\linewidth,angle=0]{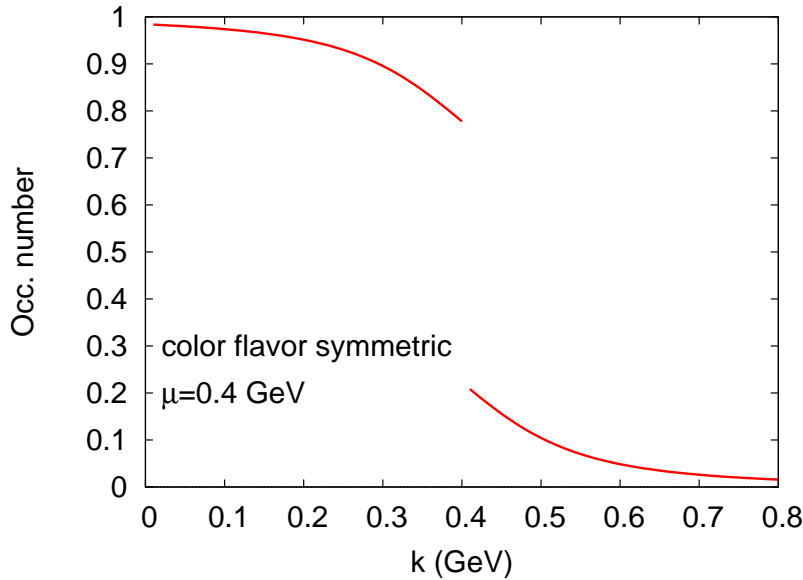} 
\caption{Occupation factors for the color-flavor symmetric
 phase. They are independent of the color-flavor combination. 
For $\mu=0.4$ GeV, $\Delta=0.0532$ GeV. The color-flavor symmetric
 phase is stable for $\mu\leq0.505$ GeV.} \label{exci3}
\end{figure}
\subsection{Color and flavor asymmetric phase}
In the case of the color and flavor asymmetric phase, for
$\epsilon_p=\sqrt{p^2+M^2}-\mu\geq 0$, the BCS vacuum is annihilated
by the opertors
$$d_{1up}=c_{1up}+K_p c^\dagger_{2d\overline p}
, \quad d_{1dp}=c_{1dp}-K_p
c^\dagger_{3u\overline p},
$$
$$
d_{2up}=c_{2up}-K_p
c^\dagger_{1d\overline p},\quad
d_{2dp}=c_{2dp}+K_pc^\dagger_{3u\overline p}
$$
$$d_{3up} =c_{3up}
,\quad d_{3dp}=c_{3dp}
,\quad d_{isp}=c_{isp},\quad i\in\{1,2,3\}$$ while for
$\epsilon_p=\sqrt{p^2+M^2}-\mu\leq 0$, it is annihilated by
$$d_{1up}=c_{1up}^\dagger-\widetilde K_pc_{2d\overline p}
,\quad d_{1dp}=c_{1dp}^\dagger-\widetilde K_pc_{2u\overline p}
$$
$$
d_{2up}=c_{2up}^\dagger+\widetilde K_p
c_{1d\overline p},\quad d_{2dp}=c_{2dp}^\dagger+\widetilde
K_p
c_{1u\overline p}
$$
$$
d_{3up}=c_{3up}^\dagger
,\quad
d_{3dp}=c_{3dp}^\dagger
,\quad d_{isp}=c_{isp}^\dagger,\quad i\in\{1,2,3\}.$$

The gap equation reads
$$1={2G\over V}\left(\sum_{p,\epsilon_p\leq0}+\sum_{p,\epsilon_p>0}\right){1\over\sqrt{\epsilon_p^2+\Delta^2}}.$$
and the thermodynamical potential is given by
\begin{eqnarray*}&&\langle \hat
K_{MFA}\rangle=\sum_{p,\epsilon_p\leq0}\left(7\epsilon_p-2\sqrt{\epsilon_p^2+\Delta^2}
\right)+\sum_{p,\epsilon_p>0}\left(2\epsilon_p-2\sqrt{\epsilon_p^2+\Delta^2}
\right)+V{\Delta^2\over2G_c}.
\end{eqnarray*}
In the color and flavor asymmetric phase, $\Delta_{eff}=\Delta/{3}.$

If written in terms of $\Delta_{eff}, $ the thermodynamical
potentials $\langle \hat K_{MFA}\rangle $ for the color-flavor
symmetric phase, for the color symmetric flavor asymmetric phase and
for the color-flavor asymmetric phase, coincide. These phases  are
essentially equivalent. The only difference is that in the
color-flavor symmetric phase the occupation numbers are color-flavor
independent and in the remaining phases they are not. The respective
BCS vacua are related to each other by simple color-flavor
rotations. To see that the color-flavor asymmetric phase is
unitarily similar to the color-symmetric flavor-asymmetric phase,
consider the matrix in $SU(\overline 3)$ which transforms the vector
$(1,0,0)$ into
$({1\over\sqrt{3}},{1\over\sqrt{3}},{1\over\sqrt{3}})$. The operator
induced by the same matrix transforms the color-flavor asymmetric
BCS vacuum into the color-symmetric flavor-asymmetric BCS vacuum.
The 2SC phase which is described in the literature (e. g., see
\cite{ruster}) differs from the color-flavor asymmetric phase here
considered in using or not using the Lagrange multiplier $\mu_8$ to
implement the constraint $\langle\lambda_8\rangle=0.$ The present
work shows that this implementation should not be imposed. Although
$n_1=n_2\neq n_3$, the color rotation mentioned above transforms the
colors $1,2,3$ into new colors $1',2',3'$ such that
$n_{1'}=n_{2'}=n_{3'}$. Nothing prevents us from regarding
$1',2',3'$ as the physical colors and $1,2,3$ as auxiliary colors.
Avoiding the constraint $\langle\lambda_8\rangle=0$ will have an
effect on the critical point which is not irrelevant because, as
fig. \ref{exci4} shows, the $P,\mu$ curves have very close slopes.

\begin{figure}[ht]
\vspace{1.5cm} \centering
\includegraphics[width=0.75\linewidth,angle=0]{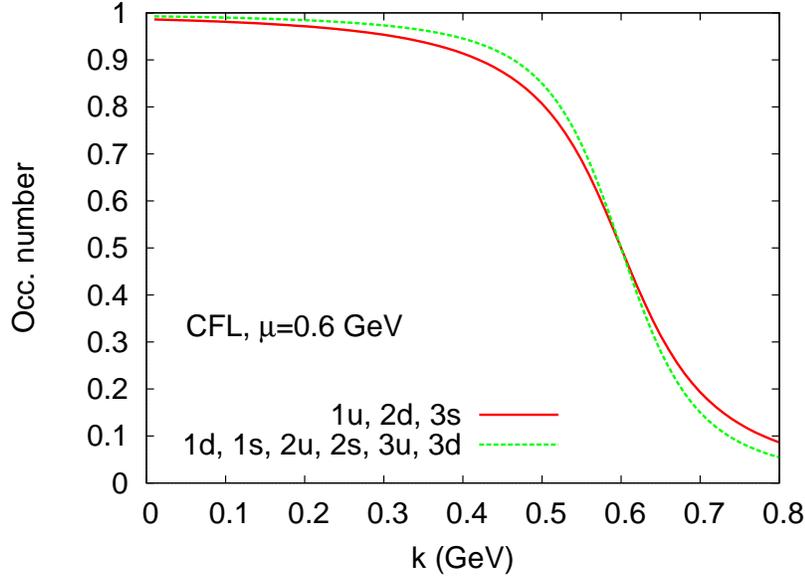} 
\caption{Occupation factors for the CFL phase. Full
curve: (1u), (2d), (3s) color-flavor combinations. Dashed curve:
remaining combinations. For $M=0$, $\mu=0.6$ GeV, $\Delta=0.102$
GeV.  The CFL phase is stable for $\mu\geq0.505$ GeV.} \label{exci2}
\end{figure}

\section{Occupation numbers}
It is clear that the curve `pressure versus chemical potential' is
the same for the phases color-flavor symmetric, color symmetric
flavor asymmetric and color-flavor asymmetric. Only the color-flavor
dependence of the occupation numbers distinguishes these phases.

 For the color-flavor symmetric phase, we find:
$$\langle\Phi|c^\dagger_{iap}c_{iap}|\Phi\rangle={7\over9}-{2\over9}~{\epsilon_p\over\sqrt{\epsilon_p^2+9\Delta^2}}\quad{\rm for}\quad
\epsilon_p\leq0,$$
$$\langle\Phi|c^\dagger_{iap}c_{iap}|\Phi\rangle={2\over9}-{2\over9}~{\epsilon_p\over\sqrt{\epsilon_p^2+9\Delta^2}}\quad{\rm for}\quad
\epsilon_p>0.$$ In Fig. \ref{exci3},  these occupation numbers are
presented.

For the color symmetric flavor asymmetric phase, if
$i\in\{1,2,3\},a\in\{u,d\}$, we find:
$$\langle\Phi|c^\dagger_{iap}c_{iap}|\Phi\rangle={2\over3}-{1\over3}~{\epsilon_p\over\sqrt{\epsilon_p^2+3\Delta^2}}\quad{\rm for}\quad
\epsilon_p\leq0,$$
$$\langle\Phi|c^\dagger_{iap}c_{iap}|\Phi\rangle={1\over3}-{1\over3}~{\epsilon_p\over\sqrt{\epsilon_p^2+3\Delta^2}}\quad{\rm for}\quad
\epsilon_p>0,$$ while
$$\langle\Phi|c^\dagger_{isp}c_{isp}|\Phi\rangle=1\quad{\rm for}\quad
\epsilon_p\leq0,\quad
\langle\Phi|c^\dagger_{isp}c_{isp}|\Phi\rangle=0\quad{\rm for}\quad
\epsilon_p>0, \quad i\in\{1,2,3\}.$$
 For the  color-flavor asymmetric phase, if
$i\in\{1,2\},a\in\{u,d\}$, we find:
$$\langle\Phi|c^\dagger_{iap}c_{iap}|\Phi\rangle={1\over2}-{1\over2}~{\epsilon_p\over\sqrt{\epsilon_p^2+\Delta^2}}\quad{\rm for}\quad
\epsilon_p\leq0,\quad{\rm and~ for}\quad \epsilon_p>0,
$$
while,
$$\langle\Phi|c^\dagger_{isp}c_{isp}|\Phi\rangle=1\quad{\rm for}\quad
\epsilon_p\leq0,\quad
\langle\Phi|c^\dagger_{isp}c_{isp}|\Phi\rangle=0\quad{\rm for}\quad
\epsilon_p>0,\quad i\in\{1,2,3\}$$
$$\langle\Phi|c^\dagger_{3ap}c_{3ap}|\Phi\rangle=1\quad{\rm for}\quad
\epsilon_p\leq0,\quad
\langle\Phi|c^\dagger_{3ap}c_{3ap}|\Phi\rangle=0\quad{\rm for}\quad
\epsilon_p>0,\quad a\in\{u,d,s\}.$$ For the CFL phase, if
$(ia)\in\{(1u),(2d),(3s)\}$, we find:
$$\langle\Phi|c^\dagger_{iap}c_{iap}|\Phi\rangle=
{1\over2}-{1\over6}{\epsilon_p\over\sqrt{\epsilon_p^2+4\Delta^2}}
-{1\over3}{\epsilon_p\over\sqrt{\epsilon_p^2+\Delta^2}}\quad{\rm
for}\quad \epsilon_p\leq0,\quad{\rm and ~for}\quad
\epsilon_p\leq0,$$
 while, if $(ia)\not\in\{(1u),(2d),(3s)\}$,
we have:
$$\langle\Phi|c^\dagger_{iap}c_{iap}|\Phi\rangle=
{1\over2}\left(1-{}~{\epsilon_p\over\sqrt{\epsilon_p^2+\Delta^2}}\right)\quad{\rm
for}\quad \epsilon_p\leq0\quad{\rm and~ for}\quad \epsilon_p>0.$$ In
Fig. \ref{exci2},  these occupation numbers are presented.
\section{Discussion and conclusions}
We investigate, in a simple schematic model of the NJL type, the
equilibrium between the 2CS pase, which is described by the
color-flavor symmetric phase, and the CFL phase of strange quark
matter at high densities. We focus on densities for which chiral
symmetry is expected to have been restored. In the color-flavor
symmetric phase, the quark numbers are independent of the
color-flavor combination. In the CFL phase the quark numbers depend
on the color-flavor combination, that is, the number of quarks
associated with the color-flavor combinations $ur,dg,sb$ is
different from the number of quarks associated with the color flavor
combinations $ug,ub,dr,db,sr,sg$.
Our results were obtained for the coupling constant
$G_c=5$GeV$^{-2}$ and are summarized in figs. \ref{exci4},
\ref{exci1}, \ref{exci6} and \ref{exci7}. It is shown by fig.
\ref{exci4} that the color-flavor symmetric phase is stable for a
chemical potential $\mu$ below $\mu_c=0.505$ GeV, while the CFL
phase is stable above, the equilibrium pressure being $P_c=0.003$
GeV$^4$. In other words, the color-flavor symmetrical phase is
stable for densities below the critical quark density $\rho_1=0.022$
GeV$^3${=$2.75$ fm$^{-3}$, which is 5.39 times the nuclear
saturation density}. The CFL phase is stable above the critical
density $\rho_2=0.025$ GeV$^3${=$3.125$ fm$^{-3}$ which is 6.13
times the nuclear saturation density}. Between $\rho_1$ and $\rho_2$
both phases co-exist. This is illustrated by fig. \ref{exci1},
which also shows that the CFL phase leads to a considerably higher
incompressibility than the one corresponding to the unphysical
continuation of the color flavor symmetric curve. We have used a
3-momentum regularizing cutoff $\Lambda=0.8$ GeV. As is shown by the
tails of the occupation factors (see figs. \ref{exci3},
\ref{exci2}), this cutoff is adequate if the relevant chemical
potential does not exceed 0.6 GeV.  
The chemical potential $\mu_c$ and the density $\rho_2$ are high
enough to qualitatively support the approximation of neglecting the
current mass of the strange quark. However, it would be interesting
to investigate within the framework of the present approach the
equilibrium between the color symmetric flavor asymmetric phase and
the CFL phase, taking into account the strange quark mass. The
present results are consistent with the hypothesis advanced in
\cite{alford2} that the interior of neutron stars may be made of
strange matter. This calculation has been performed for the
temperature $T=0$.  The 2SC phase which is described in the
literature (see, e.g. \cite{ruster}) differs from the color-flavor
asymmetric phase in that the Lagrange multiplier $\mu_8$ is used to
implement the constraint $\langle\lambda_8\rangle=0.$ The present
work shows that this procedure should not be followed, since the
color-flavor asymmetric phase is unitarily similar to the
color-flavor symmetric phase and to the color symmetric flavor
asymmetric phase and by implementing the constraint
$\langle\lambda_8\rangle=0$ the thermodynamical potential is
increased and the pressure is lowered so that the stability is
reduced.  Avoiding the constraint $\langle\lambda_8\rangle=0$ will
have an effect on the critical point which is not irrelevant
because, as fig. \ref{exci4} shows, the $P,\mu$ curves have very
close slopes.

It is important to take finite temperatures into account and we may
ask which phase prevails at high temperature: CFL or the
color-flavor symmetric one?

\section*{Acknowledgements} The present research was
partially supported by Project PTDC/Fis/113292/2009.

\end{document}